\documentclass[10pt,a4paper,twocolumn,superscriptaddress]{revtex4}

\usepackage{textcomp}
\usepackage{color}
\usepackage{amsmath}
\usepackage{amssymb}
\usepackage{mathrsfs}
\usepackage{graphicx}
\usepackage{subfig}
\usepackage{epsfig}
\usepackage{physics}
\usepackage{pdfpages}
\usepackage{hyperref}

\usepackage[squaren,Gray]{SIunits}

\begin{document}

\title{Bloch Oscillations of Driven Dissipative Solitons in a Synthetic Dimension}

\author{Nicolas Englebert}
\email{nicolas.englebert@ulb.be}
\affiliation{Service OPERA-\textit{Photonique}, Universit\'e libre de Bruxelles (U.L.B.), 50~Avenue F. D. Roosevelt, CP 194/5, B-1050 Brussels, Belgium}

\author{Nathan Goldman}
\affiliation{CENOLI, Universit\'e libre de Bruxelles (U.L.B.), CP 231, Campus Plaine, B-1050 Brussels, Belgium}

\author{Miro Erkintalo}
\affiliation{Department of Physics, University of Auckland, Auckland 1010, New Zealand}
\affiliation{The Dodd-Walls Centre for Photonic and Quantum Technologies, New Zealand}

\author{Nader Mostaan}
\affiliation{CENOLI, Universit\'e libre de Bruxelles (U.L.B.), CP 231, Campus Plaine, B-1050 Brussels, Belgium}
\affiliation{Department of Physics and Arnold Sommerfeld Center for Theoretical Physics (ASC), Ludwig-Maximilians-Universität München, Theresienstr. 37, D-80333 München, Germany}
\affiliation{Munich Center for Quantum Science and Technology (MCQST), Schellingstr. 4, D-80799 München, Germany}

\author{Simon-Pierre Gorza}
\affiliation{Service OPERA-\textit{Photonique}, Universit\'e libre de Bruxelles (U.L.B.), 50~Avenue F. D. Roosevelt, CP 194/5, B-1050 Brussels, Belgium}

\author{Fran\c{c}ois Leo}
\affiliation{Service OPERA-\textit{Photonique}, Universit\'e libre de Bruxelles (U.L.B.), 50~Avenue F. D. Roosevelt, CP 194/5, B-1050 Brussels, Belgium}

\author{Julien Fatome}
\affiliation{Department of Physics, University of Auckland, Auckland 1010, New Zealand}
\affiliation{The Dodd-Walls Centre for Photonic and Quantum Technologies, New Zealand}
\affiliation{Laboratoire Interdisciplinaire Carnot de Bourgogne, UMR 6303 CNRS ‒ Université Bourgogne-Franche-Comté, Dijon, France}

\begin{abstract}
\textbf{
The engineering of synthetic dimensions allows for the construction of fictitious lattice structures by coupling the discrete degrees of freedom of a physical system, such as the quantized modes of an electromagnetic cavity or the internal states of an atom. This method enables the study of static and dynamical Bloch band properties in the absence of a real periodic lattice structure. So far, the vast majority of implementations have focused on linear and conservative processes, with the potentially rich physics and opportunities offered by nonlinearities and dissipation remaining largely unexplored. Here, we theoretically and experimentally investigate the complex interplay between Bloch band transport, nonlinearity, and dissipation, exploring how a synthetic dimension realised in the frequency space of a coherently-driven optical resonator influences the dynamics of nonlinear waves of the system. In particular, we observe and study nonlinear dissipative Bloch oscillations occurring along the synthetic frequency dimension, sustained by localized dissipative structures (solitons) that persist endlessly in the resonator. The unique properties of the dissipative soliton states can extend the effective size of the synthetic dimension far beyond that achieved in the linear regime, as well as enable long-lived Bloch oscillations and high-resolution probing of the underlying band structure. Besides representing the first experimental study of the interplay between Bloch oscillations and dissipative solitons, our work establishes Kerr resonators  as an ideal platform for the study of nonlinear dynamics in long-scale synthetic dimensions, with promising applications in topological photonics.}  

\end{abstract}

\maketitle

\begin{figure*}
    \centering
    \includegraphics[scale=0.9]{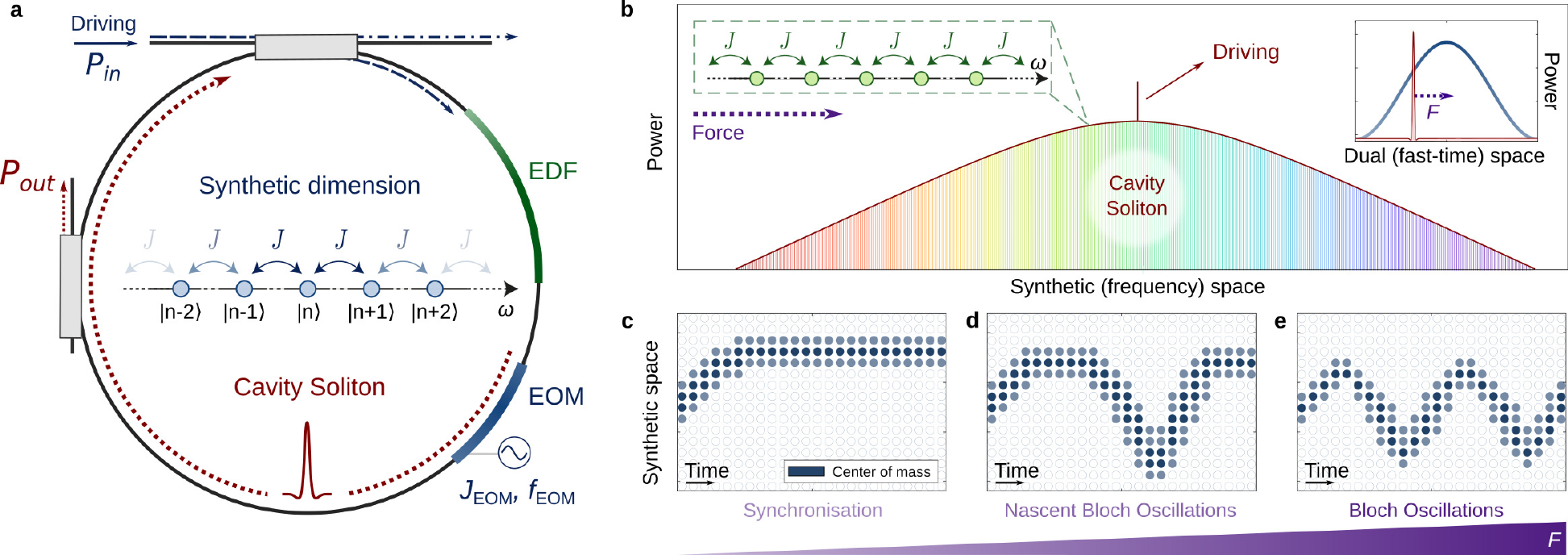}
    \caption{\textbf{Concept.}  \textbf{a}, Schematic illustration of the experimental implementation, consisting of a coherently-driven optical ring resonator with a Kerr-type nonlinearity that incorporates an electro-optic phase modulator (EOM) to realise coupling between different cavity modes. The EOM engenders hopping between resonator modes separated by an integer number of the cavity free-spectral range (FSR)
    , thus yielding a tight-binding lattice in the synthetic (frequency) space. An erbium-doped fibre amplifier (EDF) is also inserted into the cavity to compensate for the losses induced by the EOM. \textbf{b}, In the nonlinear regime, a temporal CS is generated within the resonator. The soliton is associated with a broad coherent optical frequency comb, thus extending the size of the synthetic frequency lattice with numerous equally-spaced sites. The corresponding band energy and intensity profile of the CS along the Brillouin zone are illustrated in the top-right inset. \textbf{c-e}, A constant force $F$ can be introduced along the synthetic frequency dimension by detuning the frequency of the intra-cavity EOM from an integer multiple of the cavity FSR, thus forcing the CS to undergo transport along the synthetic lattice; as the effective force $F$  increases, the soliton's transport changes from synchronisation to ideal Bloch oscillations.}
    \label{fig:Concept}
\end{figure*}

Accelerating a quantum particle on an ideal periodic lattice potential leads to oscillatory motion, instead of a net drift. This counter-intuitive phenomenon, known as Bloch oscillations (BOs), has been observed in various contexts ranging from solid-state experiments~\cite{leo1998interband} to synthetic lattice systems involving matters waves~\cite{dahan1996bloch} or light~\cite{morandotti1999experimental,bersch2009experimental,chen_real-time_2021}. Besides enabling direct observations of BOs (and other phenomena~\cite{szameit_discrete_2010,bloch2012quantum,ozawa_topological_2019}), the efficient control and agility characteristic of synthetic lattices offer a route for judiciously exploring how effects such as nonlinearities, dissipation, and Bloch-band properties influence the oscillation dynamics. While BOs can survive in a dissipative lattice~\cite{efremidis2004bloch}, the effects of nonlinearities are known to be severely detrimental~\cite{morandotti1999experimental, trombettoni2001discrete,konotop2002modulational,machholm2003band,fallani2004observation}, giving rise to instabilities that prohibit persistent oscillations. Even though methods have been proposed to prevent such instabilities~\cite{salerno2008long,gaul2009stable}, the observation of long-lived BOs is still lacking in the realm of nonlinear waves. Furthermore, the impact of dissipation on the BOs of nonlinear waves has remained largely unexplored~\cite{longstaff2020bloch}.

Beyond fundamental questions regarding their existence and stability, BOs constitute a practical probe for engineered band structures. For instance, driving BOs on a one-dimensional lattice directly reflects the bandwidth and Brillouin zone of the underlying band structure. Moreover, BOs can be used to extract geometric and topological properties of two-dimensional systems, such as the Berry curvature~\cite{price2012mapping,wimmer2017experimental,wintersperger2020realization}, quantized Wilson loops~\cite{li2016bloch,hoeller2018topological,di2020non} and the winding numbers of Floquet phases~\cite{wintersperger2020realization}. As such, exploring the specific features of BOs in the presence of nonlinearity and dissipation is crucial to elucidating the properties of novel systems where such effects play a key role. 

Designing real (physical) periodic structures for photons and matter waves constitutes a natural path towards the realisation of synthetic lattices and quantum simulators~\cite{bloch2012quantum,ozawa_topological_2019}. Interestingly, an alternative method has recently emerged by which motion is engineered along a synthetic dimension formed through coupling of the discrete degrees of freedom of a system~\cite{ozawa_topological_2019-1}. First proposed and realised in ultracold gases~\cite{celi2014synthetic,mancini2015observation,stuhl2015visualizing,an2017direct,chalopin2020probing}, the concept of synthetic dimensions has more recently entered the realm of photonics~\cite{yuan_synthetic_2018,lustig2021topological}. In particular, pioneering experiments~\cite{dutt_experimental_2019} have shown that different frequency modes of an electromagnetic cavity can be used to represent an array of fictitious lattice sites, with motion along this lattice generated by an electro-optical modulator (EOM) inserted inside the cavity  [see Fig.~\ref{fig:Concept}a]. Such a setting was recently exploited to realise artificial gauge fields for light~\cite{dutt2020single,balvcytis2021synthetic} and to reveal topological windings of non-hermitian bands~\cite{wang2021generating}. However, these experiments have so far focused on the linear operation regime, where group-velocity dispersion (and other experimental constraints) strongly limit the extent of the synthetic array. This despite the fact that the particular  (fibre-optic) cavity configurations used in these experiments come with an intrinsic (Kerr-type) nonlinearity that introduces non-local coupling between the fictitious lattice sites~\cite{yuan_creating_2020}. 

Concurrently, the nonlinear dynamics of coherently-driven Kerr-type resonators have been extensively studied in the absence of an intra-cavity EOM (and hence without a synthetic frequency dimension)\,\cite{haelterman_dissipative_1992,coen_modulational_1997,coen_continuous-wave_2001,leo_temporal_2010, herr_temporal_2014,jang_temporal_2015, pasquazi_micro-combs_2018,nielsen_nonlinear_2021, xu_spontaneous_2021,englebert_parametrically_2021,englebert_temporal_2021,erkintalo_phase_2021}. It is well-established that such devices can host a number of universal nonlinear phenomena~\cite{haelterman_dissipative_1992,coen_modulational_1997,coen_continuous-wave_2001,xu_spontaneous_2021}, amongst which the emergence of localized dissipative structures known as temporal cavity solitons (CSs) has attracted particular attention. First observed in an optical fibre ring resonator~\cite{leo_temporal_2010}, temporal CSs have been extensively studied in the context of monolithic microresonators, where they have been shown to underpin the formation of broad and coherent optical frequency combs~\cite{delhaye_optical_2007,coen_universal_2013,chembo_spatiotemporal_2013,pasquazi_micro-combs_2018}. Interestingly, whilst CSs and resonator-based synthetic dimensions have been studied in similar platforms, there is little overlap between the two domains of research. It is only very recently that a theoretical study explored how CSs behave in the presence of an intra-cavity EOM implementing a synthetic frequency dimension~\cite{tusnin_nonlinear_2020}. However, the difficulty to include an intra-cavity EOM while maintaining low intra-cavity losses, a prerequisite for CS existence, has so far prevented experimental investigations.

Here, by leveraging a recently-developed active cavity scheme\,\cite{englebert_temporal_2021} to compensate for the losses incurred by an intra-cavity EOM, we experimentally explore the nonlinear dynamics of dissipative Kerr CSs in a synthetic frequency dimension, thus bridging the gap between the realms of synthetic photonic lattices and nonlinear Kerr resonators. In addition to exploring the predictions of ref.~\cite{tusnin_nonlinear_2020}, we implement an effective force along the synthetic lattice and demonstrate that CSs can undergo long-lived BOs whose specific features can be substantially influenced by the presence of dissipation and nonlinearity. Our results establish Kerr resonators as an ideal platform to explore Bloch band transport in the presence of nonlinearity and dissipation, 
with potential implications for optical frequency combs and topological photonics~\cite{ozawa_topological_2019,ozawa_topological_2019-1,smirnova2020nonlinear}.\\

\noindent\textbf{Principle and theoretical description}\\
We begin by describing the analogy (and the limitations of the analogy) between a quantum particle on a periodic lattice and a coherently-driven Kerr resonator~\cite{dutt_experimental_2019,leo_temporal_2010}. To this end, we consider a ring resonator with an intra-cavity EOM that imparts a sinusoidal phase modulation (with amplitude $J_\mathrm{EOM}$) on the field circulating inside the resonator, with the modulation frequency $f_\mathrm{EOM}$ set close to an integer multiple of the resonance frequency spacing of the cavity (the free-spectral range, FSR): $f_\mathrm{EOM}= m\times \text{FSR} + \Delta f$, with $m\in \mathbb{N}$ and $\Delta f\ll \text{FSR}$ [Fig.~\ref{fig:Concept}a]. As described in~\cite{yuan_synthetic_2018}, the EOM permits an excitation in the $n^{th}$ cavity mode to effectively hop into nearby modes $n\pm m$ along the frequency axis, thus mimicking a tight-binding model of a quantum particle moving along a one-dimensional lattice. Therefore, under conditions of negligible nonlinearity (low intra-cavity power), weak dispersion, and balance between loss and driving, this optical system can be described by a linear Schr\"odinger equation (with $\hbar\!=\!1$) associated with the following conservative tight-binding Hamiltonian [Methods]:
\begin{align}
    \hat{H} &= -\displaystyle 
     J\sum_n(\dyad{n}{n-1}+\dyad{n}{n+1})+(F d)\sum_n n\dyad{n}{n} \label{eq:H1} \\
     & = -\phantom{J}\sum_\tau  \dyad{\tau}{\tau}\,\left\{2J \cos(d(\tau-Ft))\right\}, \label{eq:H2}
\end{align}
where the states are labelled such that $|n\pm 1\rangle$ correspond to the cavity modes $n\pm m$ coupled by the EOM. The first term of Eq.~\eqref{eq:H1} describes hopping 
between neighbouring sites $|n\rangle\rightarrow |n \pm1\rangle$ of the lattice with hopping amplitude $J=J_\mathrm{EOM}\times\text{FSR}/2$, whereas the second term describes the action of an additional constant force  $F = \Delta f/f_\mathrm{EOM}
\approx \Delta f/(m\times\text{FSR})$ along the lattice, with $d=2\pi\times m\times \text{FSR}\equiv\Omega_m$ an effective inter-site distance. Equation~\eqref{eq:H2} corresponds to the dual-space representation of the Hamiltonian: the ``momentum" coordinate $\tau$ belongs to the first Brillouin zone $[- \pi/d, \pi/d]$ and can be understood as the \textit{fast-time} coordinate along a single cavity roundtrip. (In contrast, $t$ corresponds to a \textit{slow-time} variable describing the evolution of the intra-cavity field over consecutive roundtrips~\cite{leo_temporal_2010}.) Note also that in Eq. (2), the action of the force enters through a time-dependent gauge potential defined by $Ft$. 

When $F\!=\!0$, one identifies the dispersion relation $E(\tau)\!=\!-2J\cos(d\tau)$, which reveals the existence of a band structure [top-right inset of Fig.~\ref{fig:Concept}b, in blue]. Activating the force, $F\!\ne\!0$, generates a constant drift along the momentum ($\tau$) space, hence leading to transport:~a localized wave-packet experiences an oscillating band velocity $dE/d\tau$, causing the packet to oscillate back and forth along the synthetic frequency lattice, i.e., undergo BOs. In this \textit{ideal} linear framework, the oscillatory motion is characterised by a Bloch period $T_\mathrm{BO}$ and an oscillation amplitude $A_\mathrm{BO}$ in the synthetic (frequency) space given by~\cite{dahan1996bloch}~:
 \begin{equation}T_\mathrm{BO}=\frac{2\pi}{Fd}\quad, \qquad\qquad   A_\mathrm{BO}=\frac{2J}{F}.
\label{eq:Bloch_lin}
\end{equation}

The Schrödinger equation associated with Eq.~\eqref{eq:H1} is useful in revealing the analogy between a quantum particle on a periodic lattice and an optical resonator with an intra-cavity EOM. But to permit an accurate description of the system under general conditions, the equation must be rigorously augmented with terms that take into account the full effects of nonlinearity, group-velocity dispersion, driving, and dissipation. This yields a driven-dissipative nonlinear Schrödinger equation, also known as Lugiato-Lefever equation (LLE) \cite{lugiato_spatial_1987,haelterman_dissipative_1992}, that is generalised here to account for the intra-cavity phase modulation [see Eq.\eqref{eq:LLE} in Methods]. CSs correspond to steady-state (invariant along $t$) solutions of the LLE that are localized along the fast-time ($\tau$) coordinate and characterised by a broad comb-like spectrum in the frequency ($\omega$) domain [Fig.~\ref{fig:Concept}b]. They have been extensively studied in experiments in the absence of a synthetic frequency dimension ($J_\mathrm{EOM} = 0$)\,\cite{haelterman_dissipative_1992,coen_modulational_1997,coen_continuous-wave_2001,leo_temporal_2010, herr_temporal_2014,jang_temporal_2015, pasquazi_micro-combs_2018,nielsen_nonlinear_2021, xu_spontaneous_2021,englebert_parametrically_2021,englebert_temporal_2021,erkintalo_phase_2021}. In what follows, we will present the first experiments that probe their behaviour when $J_\mathrm{EOM}\neq0$, unveiling Bloch band transport phenomena whose precise features are shaped by the interplay of nonlinearity and dissipation [see Figs.~\ref{fig:Concept}c--e]. \\

\begin{figure*}
    \centering
    \includegraphics[scale=1.05]{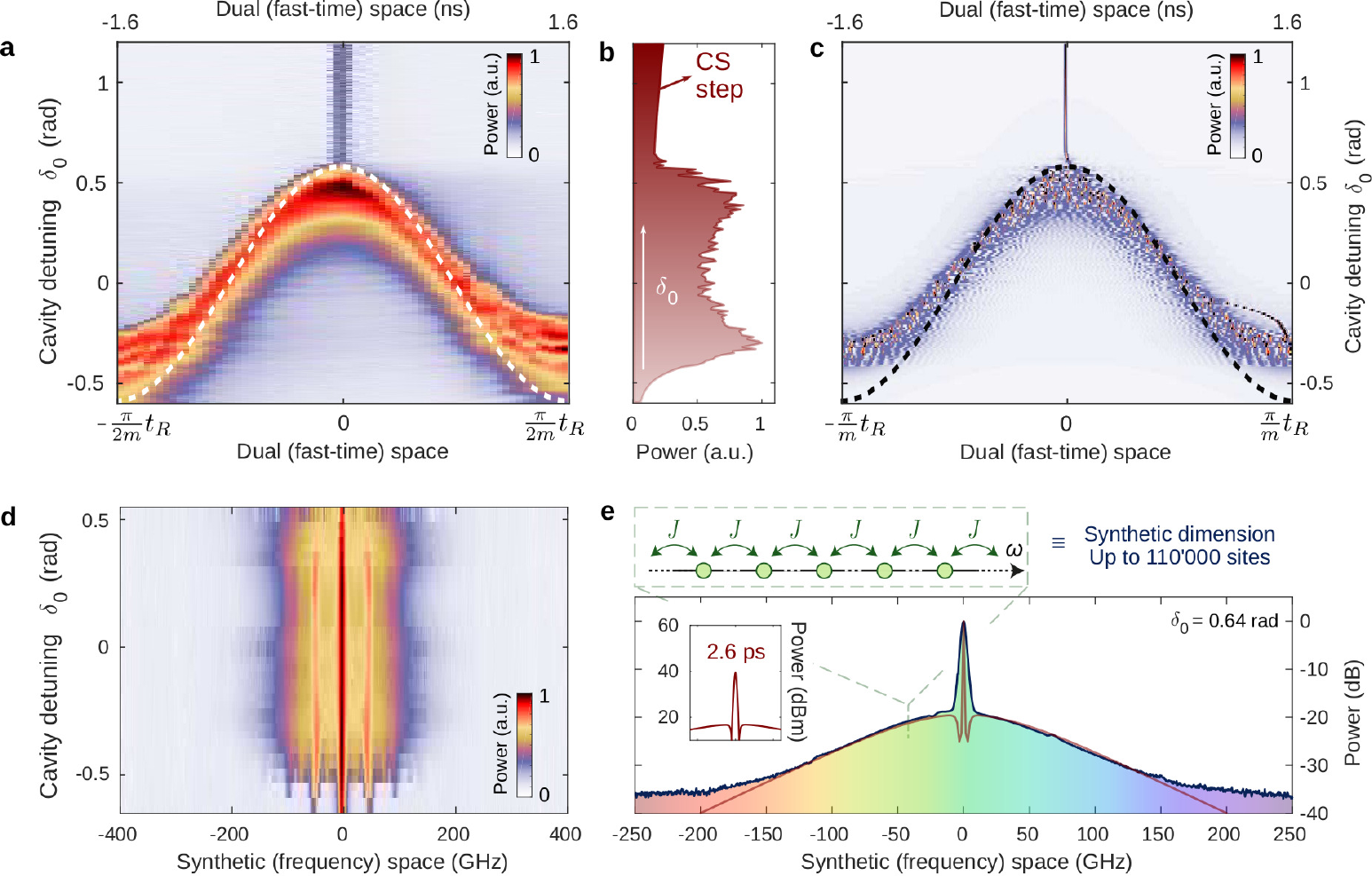}
    \caption{\textbf{Synthetic band spectroscopy in the nonlinear regime.}  \textbf{a}, Experimental measurements, showing the resonator response in the dual (fast-time) space when the mean detuning $\delta_0$ is changed with $F = 0$, $\Omega_m=2\pi\times 0.3~\mathrm{GHz}$, $J_\mathrm{EOM} = 0.6~\mathrm{rad}$ and $P_\mathrm{in}=280~\mathrm{mW}$ (for other parameters, see Methods). \textbf{b}, Evolution of the average intra-cavity intensity corresponding to \textbf{a}, showing a ``step'' feature characteristic of CS formation~\cite{herr_temporal_2014}. \textbf{c},\,Simulation results corresponding to the experiments shown in $\textbf{a}$. The dashed curves in \textbf{a} and \textbf{c} depict the linear band structure $\delta_0 = J_\mathrm{EOM}\cos(\Omega_m\tau)$ associated with the tight-binding Hamiltonian described by Eq.~\eqref{eq:H1}, and $t_\mathrm{R}=\text{FSR}^{-1}$ is the time it takes for light at the driving frequency to complete one roundtrip. \textbf{d},\,Concatenation of optical spectra measured for different cavity detunings as indicated, showing the characteristic signatures of modulational instability. \textbf{e}, Corresponding CS spectrum measured for a detuning $\delta_0 = 0.64~\mathrm{rad}$. The red solid-line indicates numerical predictions obtained from the LLE (see the modelling section in Methods), while the inset shows the corresponding simulated temporal profile of the soliton.}
    \label{fig:Soliton}
\end{figure*}

\noindent\textbf{Dissipative solitons on a synthetic band structure}\\
Our experiment consists of a 64-m-long ring resonator (yielding an FSR of 3.12~MHz) that is mainly built from standard single-mode optical fibre  [Fig.~\ref{fig:Concept}a and Methods]. The resonator is coherently-driven at 1550~nm using nanosecond, flat-top laser pulses that are synchronised to the FSR of the cavity; the driving pulses are derived from a narrow-linewidth continuous wave laser and launched into the resonator via a 99/1 coupler. 

The synthetic frequency dimension is realised by an intra-cavity EOM that is driven with a radio-frequency signal whose amplitude ($J_\mathrm{EOM}$) and frequency ($f_\mathrm{EOM}$) can be freely adjusted. Because the EOM introduces considerable losses, which hinders the accumulation of intra-cavity power levels sufficient for the observation of nonlinear effects, an erbium-doped fibre amplifier (EDF) is inserted into the cavity. This amplifier partially compensates for the loss due to the EOM~\cite{englebert_temporal_2021}, thus preserving low effective loss [Methods]. 

The band structure associated with the synthetic frequency dimension can be probed by scanning the mean phase detuning, $\delta_0$, between the driving laser and a cavity resonance -- achieved experimentally by sweeping the frequency of the laser that drives the resonator -- whilst simultaneously measuring the evolution of the intra-cavity intensity profile along the fast-time domain (see ref.~\cite{dutt_experimental_2019} and Methods). Figure~\ref{fig:Soliton}a shows results from such experiments when the EOM frequency is perfectly synchronised with an integer multiple of the FSR ($\Delta f = 0\Rightarrow F=0$). Here, the two-dimensional pseudocolor plot was obtained by vertically concatenating individual oscilloscope traces measured at the output port of the 90/10 coupler as the mean detuning $\delta_0$ is linearly increased. 

Similarly to when operating in the linear regime~\cite{dutt_experimental_2019}, the phase modulation induced by the intra-cavity EOM engenders a dynamical modification of the cavity resonance condition along the fast-time domain, thus bending the resonator’s response along the dual space. The dashed curve in Fig.~\ref{fig:Soliton}a shows that the linear band structure defined through $\delta_0 = J_\mathrm{EOM}\cos(\Omega_m\tau)$ qualitatively follows the observed response, but with deviations due to the additional effect of nonlinearity\,\cite{tusnin_nonlinear_2020}.  A more conspicuous contrast with the linear theory is the emergence of a localized structure in dual space as $\delta_0$ increases beyond the top of the band. This structure corresponds to a CS, emerging spontaneously as the mean detuning $\delta_0$ is swept across a resonance [see also Fig.~\ref{fig:Soliton}b].

To gain more insight, Fig.~\ref{fig:Soliton}c shows results from numerical simulations of the generalised LLE with parameters corresponding to our measurements of Fig.~\ref{fig:Soliton}a [Methods]. Like the experiments, the simulations show clearly the bending of the resonator's response, as well as the emergence of the CS at the top of the band structure. But in addition, the simulations also reveal that the band itself exhibits complex internal structures, being composed of highly-localized wave-packets that fluctuate as $\delta_0$ increases. This internal structure stems from the nonlinear energy transfer between different cavity modes, manifesting itself in the dual (fast-time) space as a Turing-type modulation instability (MI) that breaks the locally quasi-homogeneous intra-cavity field into fast fluctuating wave-packets\,\cite{haelterman_dissipative_1992,coen_modulational_1997,coen_continuous-wave_2001}. Because the threshold condition of the instability depends upon the cavity detuning (that itself depends upon fast-time due to the intra-cavity EOM), this instability remains localized to within the band~\cite{tusnin_nonlinear_2020, nielsen_nonlinear_2021}. 

The picosecond-scale internal structure of the band (as well as the precise profile of the CS) is not directly resolved in our experiments [Fig.~\ref{fig:Soliton}a] due to the finite 12-GHz-bandwidth of our time-domain detection system [Methods]. However, measurements made along the synthetic (frequency) space completely corroborate the simulation results. Figure~\ref{fig:Soliton}d shows the optical spectrum of the intra-cavity field measured across a range of cavity detunings where simulations predict band-MI to manifest itself. These measurements reveal a wave-packet width in the synthetic (frequency) space of about 200~GHz, commensurate with the simulated picosecond-scale features in the dual (fast-time) space.     

Figure~\ref{fig:Soliton}e shows the experimentally measured optical spectrum corresponding to the CS state, obtained for $\delta_0 = 0.64~\mathrm{rad}$. The spectrum has a sech-squared profile with a 3-dB bandwidth of 0.125\,THz, corresponding to a 2.6-ps wave-packet in duration [inset in Fig.~\ref{fig:Soliton}e]. Because the CS circulates without distortion around the resonator, its spectrum  consists of sharp discrete lines (a frequency comb) that are equally-spaced by the resonator FSR. This equi-spacing of the CS's spectral lines can be attributed to the cancellation of group-velocity dispersion by the solitons' nonlinearity, and is in stark contrast with the linear resonator modes that are not equally-spaced due to dispersion -- a fact that fundamentally limits the extent of synthetic arrays attainable under linear operation conditions.  Compounded by its broad spectral width, with more than 111,000 lines above the $-15~\mathrm{dB}$ level, the CS state can thus be seen to substantially extend the number of frequency modes (synthetic lattice sites) that are coupled through the intra-cavity modulator.\\

\noindent\textbf{Bloch oscillations of cavity solitons}\\
To the best of our knowledge, the results shown in Fig.~\ref{fig:Soliton} represent the first experimental observation of dissipative solitons in the presence of a synthetic frequency dimension. In addition to confirming the salient theoretical descriptions of ref.~\cite{tusnin_nonlinear_2020}, these results present CSs as a promising, high-resolution tool for the study of Bloch band transport in the presence of nonlinearity and dissipation. To probe for such transport, we experimentally introduce an effective force $F$ along the synthetic frequency dimension by slightly detuning
the EOM frequency from a harmonic of the cavity FSR ($\Delta f\neq 0 \Rightarrow F\neq 0$). We then use an ultrafast dispersive Fourier transform scheme~\cite{goda_dispersive_2013,mahjoubfar_time_2017} to record the roundtrip-by-roundtrip evolution of the intra-cavity optical spectrum in a regime where a single CS circulates the cavity [see Methods].  

\begin{figure*}
    \centering
    \includegraphics[scale=0.9]{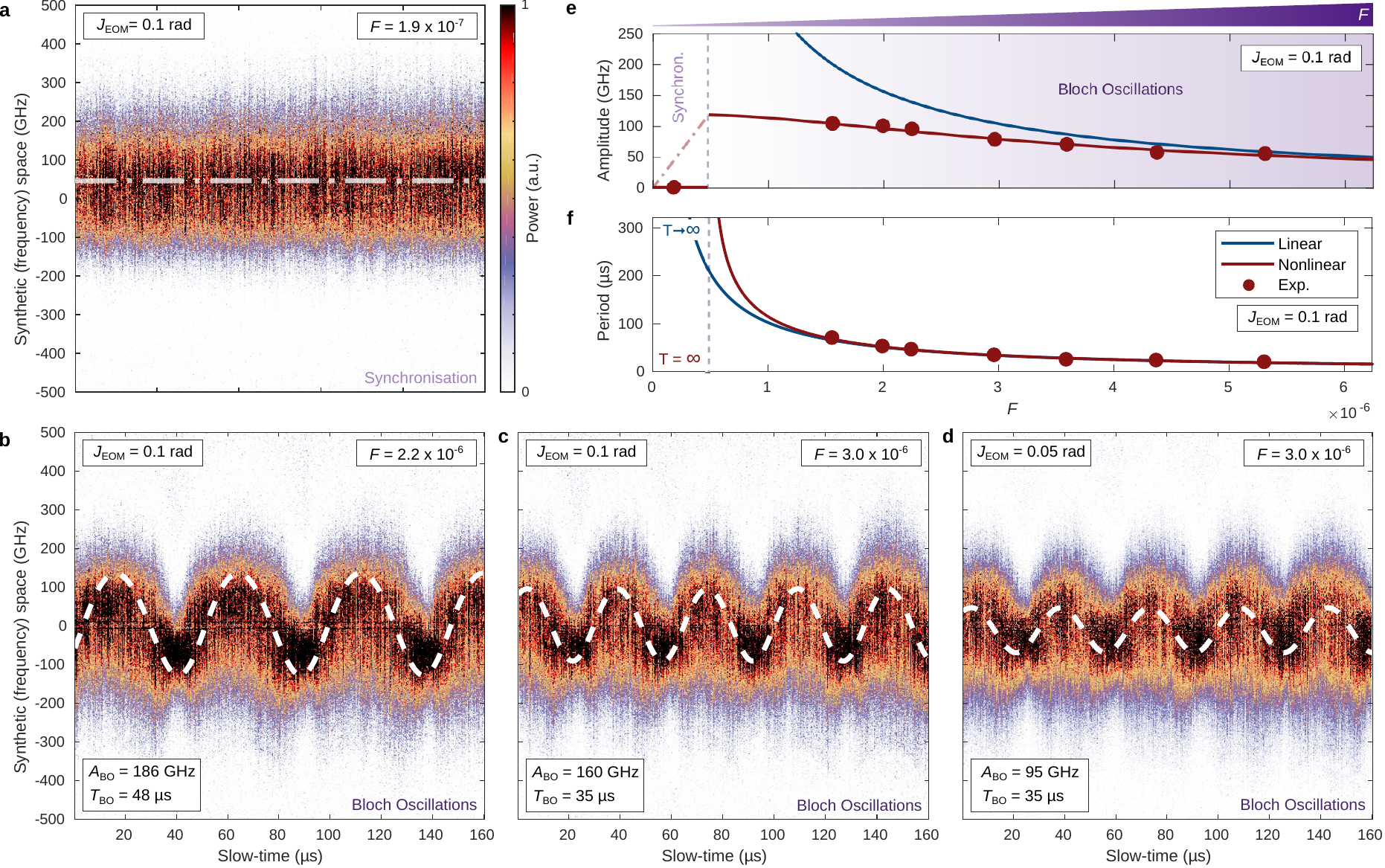}
    \caption{\textbf{Bloch oscillations of dissipative solitons.}  \textbf{a-d}, Concatenation of experimentally measured CS optical spectra as a function of time for different levels of effective force $F$ and modulation amplitude $J_\mathrm{EOM}$ (as indicated). Each measurement was obtained with $P_\mathrm{in} = 280~\mathrm{mW}$, $\Omega_\mathrm{m}=2\pi\times9.7~\mathrm{GHz}$, and $\delta_0 = 1.5~\mathrm{rad}$ (for other parameters, see Methods). White dash-dotted line in \textbf{a} shows the theoretical prediction from the nonlinear theory Eq.~\eqref{eq:Bloch_NL}, while the dashed curves in \textbf{b-d} show oscillations corresponding to ideal BOs with period and amplitude given by Eqs.~\eqref{eq:Bloch_lin} (see also main text). Red circles in \textbf{e} and \textbf{f} respectively show measured amplitude and period of CS oscillations in the synthetic frequency space as a function of the effective force $F$, while the red and blue curves show theoretical predictions from the linear [Eqs.~\eqref{eq:Bloch_lin}, blue curve] and nonlinear [Eqs.~\eqref{eq:Bloch_NL}, red curve] theories.
    The error bars reflecting the residues of the nonlinear regression to a sine are smaller than the circles.}
    \label{fig:Bloch}
\end{figure*}

Figures~\ref{fig:Bloch}a--d display typical trajectories measured in the synthetic (frequency) space for different effective force $F$. Here the pseudo-color plots show a concatenation of CS optical spectra measured on a roundtrip-by-roundtrip basis as the soliton circulates around the ring, hence illustrating the motion of a wave-packet along the synthetic lattice. These measurements immediately reveal two qualitatively different regimes. First, for small effective force $F$ [Fig.~\ref{fig:Bloch}a], the CS experiences a fast transient motion along the synthetic (frequency) dimension (not shown), but eventually comes to a halt at a (synthetic) position that is offset from the soliton's initial location (corresponding to the carrier frequency of the driving field). This behaviour is a manifestation of nonlinear \textit{synchronisation}: the modulation induced by the EOM pins the CS, shifting its mean frequency so as to cancel the temporal desynchronisation between the two (see refs.~\cite{jang_temporal_2015,cole_kerr-microresonator_2018, erkintalo_phase_2021} for related CS synchronisation involving different mechanisms).

In stark contrast, for large effective forces $F$ [Fig.~\ref{fig:Bloch}b--d], the pinning induced by the modulation is insufficient to cancel the external desynchronisation, resulting in the CS exhibiting oscillations in the synthetic (frequency) space. Physically corresponding to periodic red- and blue-shifting of the solitons' frequency, these oscillations are analogous to BOs occurring along the systems' synthetic dimension. Indeed, the dashed curves in Figs.~\ref{fig:Bloch}b--d show the sinusoidal trajectories with amplitude and period corresponding to ideal BOs given by Eqs.~\eqref{eq:Bloch_lin}. As can be seen, predictions from the ideal BOs theory are in excellent agreement with the experimental measurements, confirming that the oscillations experienced by the dissipative CSs for large effective force $F$ can be likened to BOs. \\

To better understand the transition between the two qualitatively distinct dynamical regimes observed in Figs.~\ref{fig:Bloch}a--d, we performed experiments over a wide range of effective forces $F$. Results from these experiments are summarised as red circles in Figs.~\ref{fig:Bloch}e and f, where we respectively depict the amplitude and period of the oscillations experienced by the solitons in the synthetic (frequency) space. These measurements clearly reveal how the CSs transition from synchronisation to oscillations as the effective force $F$ increases, and also how the oscillation characteristics tend towards those predicted by linear BOs theory (blue curve) as the magnitude of the effective force $F$ increases.

All of the experimental results presented in Fig.~\ref{fig:Bloch} are in excellent agreement with predictions from the LLE [Methods]. Indeed, the red curves in Figs.~\ref{fig:Bloch}e and f show oscillation characteristics extracted from a nonlinear theory derived by applying a Lagrangian perturbative approach to the CS solutions (see refs.~\cite{matsko_timing_2013,yi_theory_2016} and Supplementary Information).
This framework yields the following generalised transport equations in the synthetic (frequency) space and the dual (fast-time) domain:
\begin{align}
    &t_\mathrm{R}\dfrac{d\langle\omega\rangle}{dt}=\displaystyle-\Lambda_\mathrm{e}\langle\omega\rangle  + J_\mathrm{EOM}\Omega_m  \sin(\Omega_{m}\langle\tau\rangle) , \notag \\
    &t_\mathrm{R}\dfrac{d\langle\tau\rangle}{dt}=t_\mathrm{R}F+\beta_2L_c\langle\omega\rangle , \label{eq:Bloch_NL} 
\end{align}
where $\langle\omega\rangle$ and $\langle\tau\rangle$ represent the CS's centre of mass along the frequency and fast-time coordinates, respectively, $L_c$ is the resonator circumference, $t_\mathrm{R}=\text{FSR}^{-1}$, $\Lambda_\mathrm{e}$ is the effective resonator dissipation expressed as a proportion of intensity lost each roundtrip,  and $\beta_2$ is the group-velocity dispersion coefficient of the resonator waveguide.

The red curves in Figs.~\ref{fig:Bloch}e and f were obtained via direct numerical integration of Eqs.~\eqref{eq:Bloch_NL}, and show excellent agreement with our experiments. Additional insights can be obtained by analysing the equations in the limits where $F$ is small and large, respectively. In the first limit ($F\sim 0$), the frequency variable $\langle \omega \rangle$ evolves slowly, being slaved to the adiabatically changing temporal variable $\langle \tau \rangle$. This allows to approximate $d\langle\omega\rangle/dt\approx 0$, yielding the following Adler-like synchronisation equation~\cite{delhaye_self-injection_2014,jang_synchronization_2018,erkintalo_phase_2021} for the temporal variable $\langle \tau \rangle$ (see also Supplementary Information):
\begin{equation}
t_\mathrm{R}\frac{d\langle\tau\rangle}{dt}= Ft_\mathrm{R}+\beta_2 L_c\frac{J_\mathrm{EOM}\Omega_m}{\Lambda_\mathrm{e}}\sin(\Omega_{m}\langle\tau\rangle).
\label{eq:Adler}
\end{equation}
It should be clear that a fixed point exists for $\langle\tau\rangle$ provided that $|F|<|F_\mathrm{c}|=|\beta_2L_cJ_\mathrm{EOM}\Omega_m/(\Lambda_\mathrm{e}t_\mathrm{R})|$, in which case the CS will be synchronised to the modulation induced by the EOM, and the soliton's mean frequency correspondingly pinned to a constant value. 

In the opposite limit, $|F|t_\mathrm{R}\gg |\beta_2L_c\langle\omega\rangle|$, we have ${\langle \tau\rangle \approx Ft}$. If additionally $|F|\gg \Lambda_\mathrm{e}/(2\pi m)$, the solution to the CSs position in the synthetic (frequency) space can be written as $\langle \omega(t) \rangle \approx \langle \omega(0)\rangle + J_\mathrm{EOM}\Omega_m/(2\pi mF)\cos(\Omega_m F t)$, revealing the expected BO period $T_\mathrm{BO} = 2\pi/(\Omega_m F)$ and amplitude $A=J_\mathrm{EOM}\Omega_m/(2\pi m F)=2J/F$ in agreement with Eq.~\eqref{eq:Bloch_lin}.

The analysis above shows that the generalised transport equations~\eqref{eq:Bloch_NL} well capture the nonlinear soliton behaviour for small and large effective force $F$, and that in the latter limit, the oscillations display characteristics expected for BOs. These equations also reveal the physical mechanism that causes the oscillation characteristics to deviate from ideal BOs in the regime of intermediate force $F$: the cavity dissipation dampens the oscillations in the synthetic space, leading to an amplitude that is reduced from the conservative limit.\\

\noindent\textbf{Discussion}\\
We have reported on the generation and manipulation of  dissipative solitons in a synthetic frequency dimension. Our experiments were performed in a coherently-driven Kerr-type resonator that incorporates an EOM so as to introduce a synthetic 1D (tight-binding) lattice along the frequency axis. We then demonstrated that dissipative Kerr CSs can spontaneously emerge in the resonator when probing the systems' band structure. By detuning the EOM frequency from an integer multiple of the cavity FSR, we implemented an effective force along the synthetic frequency dimension, exploiting the CS states for the study of Bloch transport in the presence of nonlinearity and dissipation. Our experiments show that, depending on the magnitude of the effective force, the CSs can undergo synchronisation or oscillation, with the oscillation characteristics tending towards those attributed to ideal BOs in the large-force limit.

Compared to resonator-based systems operating in the linear regime, the unique characteristics of CSs enable to substantially extend the size of the synthetic 1D tight-binding lattice. Furthermore, the solitons can be used as a highly-localized tool to probe the system's Bloch band structure with unprecedented resolution: whilst direct band structure spectroscopy (see Fig.~\ref{fig:Soliton}a and ref.~\cite{dutt_experimental_2019}) can only resolve band features larger than the photo-detector response time, the amplitude of CS BOs -- which directly reflect the band structure -- is inversely proportional to the width of the Brillouin zone, thus becoming more and more discernible as the band features become smaller. In this context, it is worth noting that the width of the Brillouin zone explored by the CSs in Fig.~\ref{fig:Bloch} is an order of magnitude smaller than that of the band probed in Fig.~\ref{fig:Soliton}, and three orders of magnitude smaller than those studied in ref.~\cite{dutt_experimental_2019}. 

Several extensions of our present work can be envisioned. For example, CSs could permit to study the impact of dissipation and nonlinearity in high-dimensional synthetic frequency crystals, implemented by driving the intra-cavity EOM with multiple-tones RF signals \cite{dutt_experimental_2019}. Non-Hermitian lattices in the frequency space could be also engineered through simultaneous amplitude and phase modulations of the Kerr resonator \cite{wang2021generating}. Finally, more complex synthetic structures with non-trivial topologies could be generated by using multiple coupled or multi-component, modulated cavities~\cite{regensburger_paritytime_2012,wimmer_observation_2015-1,dutt2020single,leefmans_topological_2021,mittal_topological_2021}, whereas on-chip systems could be engineered by leveraging state-of-the-art photonic integration~\cite{hu_realization_2020,balvcytis2021synthetic}. To conclude, our work establishes that synthetic dimension engineering in Kerr-type optical resonators can be used to study how nonlinearity and dissipation affect Bloch band transport, whilst simultaneously permitting new avenues for the manipulation of dissipative CSs with potential applications in optical frequency comb generation and nonlinear topological photonics~\cite{smirnova2020nonlinear}.

\section*{Methods}

\small
\subparagraph*{\hskip-10ptModelling}\ \\
Accurate modelling of the Kerr cavity dynamics requires that the linear Schrödinger equation associated with Eq.\,\eqref{eq:H1} is rigorously augmented with terms that take into account the full effects of nonlinearity, group-velocity dispersion, driving, and dissipation. This yields the so-called Lugiato-Lefever equation (LLE) \cite{lugiato_spatial_1987,haelterman_dissipative_1992}, generalised to account for the intra-cavity phase modulation~\cite{tusnin_nonlinear_2020}:
\begin{align}
    it_\mathrm{R} \dfrac{\partial A}{\partial t} &-\dfrac{\beta_2L_c}{2}\dfrac{\partial^2A}{\partial \tau^2} +\gamma L_c|A|^2A= \label{eq:LLE} \\
    &-i\dfrac{\Lambda_\mathrm{e}}{2}A +i\sqrt{\theta P_\mathrm{in}} + [\delta_0-J_\mathrm{EOM}\cos(\Omega_m(\tau-Ft))]A. \notag
\end{align}
Here, $A(t,\tau)$ describes the electric field envelope of the intra-cavity field in the dual (fast-time) space, $L_c$ is the resonator circumference, and $t_\mathrm{R}=\text{FSR}^{-1}$ is the roundtrip time at the driving wavelength. $\beta_2$ and $\gamma$ are respectively the group-velocity dispersion and the nonlinear Kerr coefficient of the resonator waveguide, $\Lambda_\mathrm{e}$ denotes the effective cavity loss that take into account the intra-cavity gain of the optical amplifier~\cite{englebert_temporal_2021}, $\theta$ and $P_\mathrm{in}$ correspond respectively to the input coupler ratio and coherent driving power, and $\delta_0$ is the mean-value of the phase detuning between the driving field and a cavity resonance. Note that the LLE is written in a reference frame moving at the group-velocity of light at the driving wavelength. In the presence of a desynchronisation between the EOM frequency and an integer multiple of the cavity FSR ($\Delta f\neq0\Rightarrow F\neq 0$), the modulation imparted by the EOM drifts in this reference frame. On the other hand, by shifting the reference frame such that $\tau\rightarrow \tau-Ft$, we obtain a modified LLE where the modulation is stationary:
\begin{align}
    it_\mathrm{R} \dfrac{\partial A}{\partial t} &-it_\mathrm{R}F\dfrac{\partial A}{\partial \tau}-\dfrac{\beta_2L_c}{2}\dfrac{\partial^2A}{\partial \tau^2} +\gamma L_c|A|^2A= \label{LLE2} \\
    &-i\dfrac{\Lambda_\mathrm{e}}{2}A +i\sqrt{\theta P_\mathrm{in}} + [\delta_0-J_\mathrm{EOM}\cos(\Omega_m\tau)]A. \notag
\end{align}
By retaining only the first two terms (the last term) on the left-hand side (right-hand side) of Eq.~\eqref{LLE2} and applying the Fourier transformation, one obtains the linear Schrödinger equation associated with Eq.\,\eqref{eq:H1}.

The simulation results shown in Fig.~\ref{fig:Soliton} were obtained by numerically integrating Eq.~\eqref{LLE2} using the split-step Fourier method with parameters corresponding to the experiments: $L_c=64\,$m, $\gamma = 10^{-3}$\,W$^{-1}$.m$^{-1}$, $\beta_2=-20\times10^{-27}$\,s$^2$.m$^{-1}$, $\Lambda_e=0.05$, $\theta = 0.1$, $P_{in}=0.28$\,W, $J_\mathrm{EOM}=0.6$\,rad, $f_\mathrm{EOM}=0.3\,$GHz. The simulations shown in Fig.~\ref{fig:Soliton}c were obtained with a noisy initial condition and with the mean phase detuning $\delta_0$ linearly increasing at a rate of 66.5\,mrad/roundtrip. The simulations shown in Fig.~\ref{fig:Soliton}e used a hyperbolic secant initial condition and with the mean detuning fixed at $\delta_0=0.64\,$rad.

\subparagraph*{\hskip-10pt Experimental set-up}\ \\
The resonator schematically displayed in Fig.~\ref{fig:Concept}a is made of $\sim63$\,m of standard telecommunication single-mode fibre (SMF-28) spliced to a 75\,cm-long segment of erbium doped fibre (EDF, Liekki$^{\text{®}}$ ER16-8/125), leading to a cavity FSR of 3.12\,MHz. The amplifying section provides the optical gain required to partially compensate for the cavity roundtrip losses (close to 3.2 dB). The length of the amplifying section has been carefully adjusted following the method described in ref.\,\cite{englebert_temporal_2021}. The erbium doped fibre is surrounded by two wavelength division multiplexers (WDMs) to combine the 1480-nm backward pump with the intra-cavity signal, and to reject the residual non-absorbed pump power. Without this amplifying fibre segment, the total losses of the cavity have been evaluated to 3.2\,dB. A 99/1 tap-coupler is used to inject the coherent driving beam into the cavity, while part of the intra-cavity power is extracted for analysis thanks to a 90/10 coupler. Before the input coupler, a polarisation controller is used to align the polarisation state of the driving beam along one of the two principal eigenmodes of the cavity. The synthetic dimension is implemented in the frequency domain using a commercial phase modulator (EOM, bandwidth: 12\,GHz), inserted directly into the ring resonator. The EOM is driven by a radio-frequency (RF) signal generator whose amplitude ($J_\mathrm{EOM}$) and frequency ($f_\mathrm{EOM}$) can be freely adjusted. In addition, an intra-cavity polarisation controller limits the EOM polarisation-dependent losses.
The resonator is coherently-driven by means of a sub-100~Hz linewidth continuous wave (cw) laser, slightly tunable and centred at 1550.12\,nm. At this driving wavelength, the overall resonator displays anomalous group velocity dispersion -- a requisite for CS existence. The cw-laser is modulated by means of a Mach-Zehnder amplitude modulator (bandwidth: 12 GHz, extinction ratio: 30\,dB), driven by a pulse pattern generator synchronised with the cavity FSR, so as to generate a train of 1.2-ns square pulses. The resulting flat-top pulse train is then amplified with a commercial erbium doped fibre amplifier (EDFA) before injection into the ring resonator. 
We probe the band structure by slowly increasing the driving laser wavelength and recording the  power at the output of the 90/10 coupler with a 200\,kHz photodiode (\autoref{fig:Soliton}b) or with a 12\,GHz detection system (\autoref{fig:Soliton}a), as described below.
The cavity can be stabilised at any detuning $\delta_0$ by means of a counter-propagating and frequency shifted control signal~\cite{li_experimental_2020}. This signal is generated by extracting 5\% of the driving cw-laser and sending it to a tunable opto-acoustic frequency-shifter ($ 110\pm5\,$MHz). The resulting low power control signal (0.1\,mW) is then injected into the cavity on the same polarisation state as the main signal thanks to an optical circulator combined to a polarisation controller. The cavity detuning is stabilised through feedback on the driving cw-laser wavelength so as to maintain a constant intra-cavity average power for the control signal. The feedback signal is generated by a proportional-integral-derivative (PID) controller, driven by a 200\,kHz photodiode.  We then map the detuning to the frequency of the control signal by fitting the cavity transmission in the linear regime. Finally, at the output of the system, the intra-cavity field is characterised both in the temporal and spectral domains by means of a 45\,GHz photodiode connected to a 12\,GHz real-time oscilloscope (20 Gsample.s$^{-1}$) and an optical spectrum analyser, respectively. Note that a complete schematic of this experimental set-up is given in the Supplementary Information.

\subparagraph*{\hskip-10pt Bloch Oscillations}\ \\
The CS evolution along the synthetic space is recorded roundtrip-by-roundtrip by performing a dispersive Fourier transformation (DFT)\,\cite{goda_dispersive_2013,mahjoubfar_time_2017}. Using two 50\,km-long spools of standard single-mode fibre [SMF-28, $\beta_2=-20\times10^{-27}$\,s$^2$.m$^{-1}$], our real-time spectrum measurements reach a resolution of 12\,GHz\,\cite{chen_real-time_2021}. To cope with propagation losses (10\,dB/spool), each spool is surrounded by two WDMs (1450/1550\,nm) to perform (co-propagating) Raman amplification. The 1450\,nm Raman pump power, before being split in two, is set to 1.4\,W. Before the DFT, to avoid the detrimental interferences occurring between the CS and its associated cw background, a nonlinear optical loop mirror (NOLM) has been implemented to remove the cw component. The NOLM is made of a fibre loop closed on a 50/50 coupler and includes a variable optical attenuator (VOA), 200\,m of dispersion compensating fibre (DCF, nonlinearity coefficient $\gamma_{\text{DCF}} = 3\,$W$^{-1}$.km$^{-1}$) and a polarisation controller. When the VOA is set to 5\%, the NOLM transmission reaches almost one for a CS of 50\,W peak power, whereas its cw background is attenuated by 20\,dB\,\cite{smirnov_layout_2017}. A commercial EDFA is used at the input of the NOLM to increase the CS peak power above 50\,W. As the amplitude of CS BO depends on the EOM frequency, the latter is set close to $\Omega_{m} = 2\pi\times 9.7\,$GHz to be sufficiently above the DFT resolution (12\,GHz). Finally, BOs of CSs are recorded while slowly scanning the cavity detuning in the vicinity of $\delta_0=1.5$\,rad; this scan is so slow that the detuning can be considered constant over the duration of the measurements shown in Fig.~\ref{fig:Bloch}.  A complete experimental set-up is given in the Supplementary Information. \\

\textit{Note added:} We note that Bloch oscillations were recently detected along a synthetic dimension of atomic harmonic trap states\,\cite{oliver_bloch_2021}.

\section*{Acknowledgements}

\noindent This work was supported by funding from the European Research Council (ERC) under the European Union’s Horizon 2020 research and innovation programme, grant agreement No 757800 (QuadraComb) and No 716908 (TopoCold). N.E. acknowledges the support of the ``Fonds pour la formation à la Recherche dans l’Industrie et dans l’Agriculture" (FRIA-FNRS, Belgium). F.L. and N.G. acknowledge the support of the ``Fonds de la Recherche Scientifique" (FNRS, Belgium).
J.F. acknowledges the financial support from the CNRS, IRP Wall-IN project and FEDER, Optiflex project. M.E acknowledges the Marsden Fund and the Rutherford Discovery Fellowships of The Royal Society of New Zealand Te Apārangi. N. M. acknowledges funding by the Deutsche Forschungsgemeinschaft (DFG, German Research Foundation) under Germany’s Excellence Strategy – EXC-2111 – 390814868.\\

\section*{Author Contributions}

All authors contributed to the conception of the research, analysed and interpreted the results. N.E. performed the experiments and derived the reduced model with supervision from S.-P.G. and F.L. N.E., J.F. and M.E. performed simulations of the LLE and the reduced model. N.E., N.G., M.E. and J.F. prepared the manuscript, with inputs from all authors.

\section*{Data Availability}
\noindent The data that support the findings of this study are available from the corresponding author upon reasonable request.

\section*{Competing Financial Interests statement}
\noindent N.E., S.-P.G. and F.L. have filed patent applications on the active resonator design and its use for frequency conversion (European patent office, application number EP20188731.2). The remaining authors declare no competing interests. 

\bibliography{Soliton_BOs} 
\bibliographystyle{naturemag}

\newpage \ \\
\newpage
\includepdf[page=1]{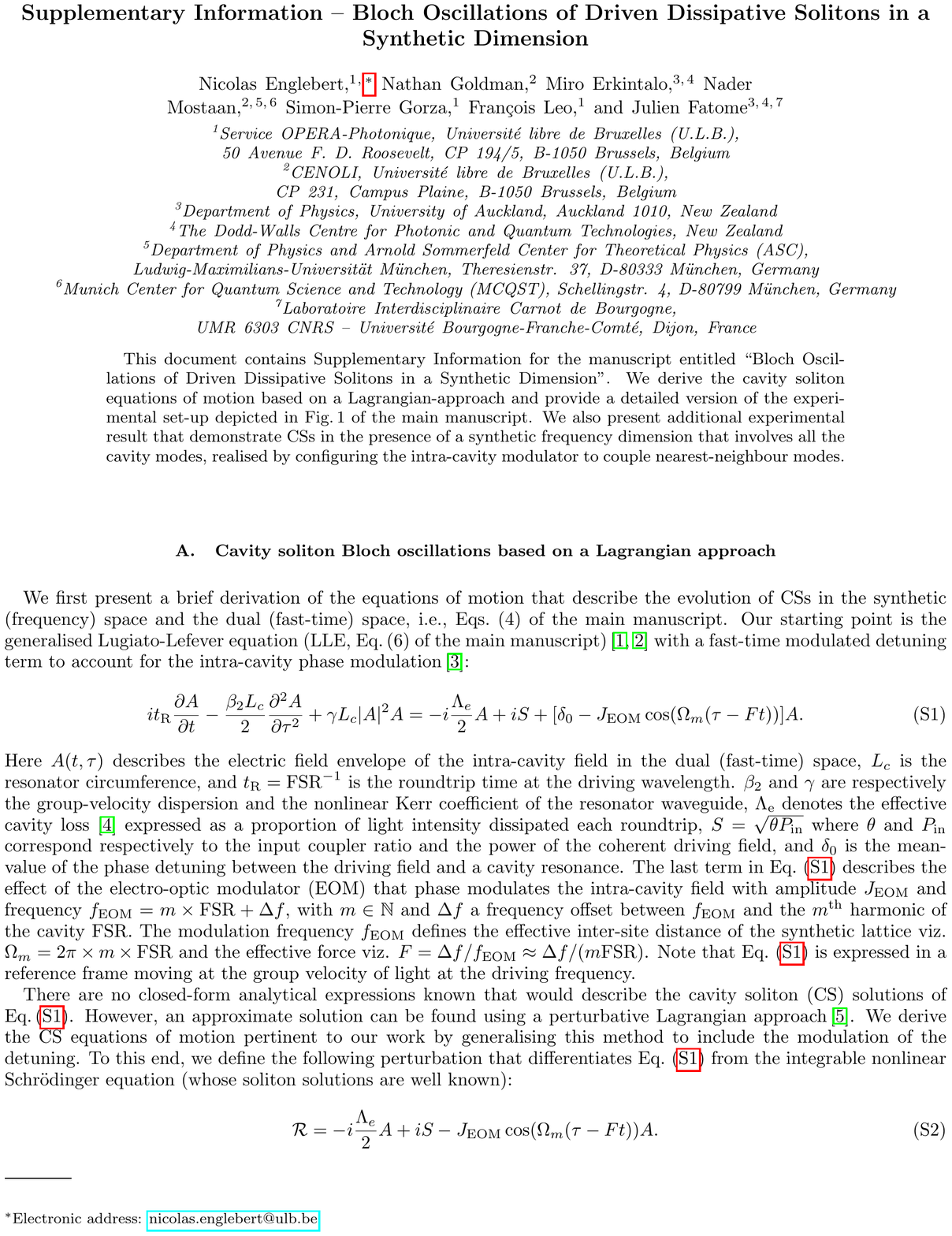} \newpage\ \newpage
\includepdf[page=2]{SI.pdf} \newpage\ \newpage
\includepdf[page=3]{SI.pdf} \newpage\ \newpage
\includepdf[page=4]{SI.pdf} \newpage\ \newpage
\includepdf[page=5]{SI.pdf} \newpage\ \newpage
\includepdf[page=6]{SI.pdf} \newpage\ \newpage
\includepdf[page=7]{SI.pdf} 
\end{document}